\documentclass[%
 reprint,
superscriptaddress,
showpacs,
showkeys,
 amsmath,amssymb,
nofootinbib
]{revtex4-1}
\usepackage{array}
\usepackage{amsmath}
\usepackage{graphicx}
\usepackage{dcolumn}
\usepackage[italian, english]{babel}
\usepackage[utf8x]{inputenc}
\usepackage{tabularx}
\usepackage{bm}
\usepackage[sort&compress]{natbib}
\usepackage{multirow}
\usepackage{longtable}
\usepackage{bigstrut}
\usepackage[para]{threeparttable}

\newcommand{\equatname}{Eq. }
\newcommand{\figname}{Fig. }
\newcommand{\tabname}{Table }
\newcommand{\secname}{Sec. }

\pdfoutput=1

\usepackage[pdftex,breaklinks,colorlinks,
citecolor=blue,
urlcolor=blue,
linkcolor=blue,
pdftitle={Reference worldwide model for antineutrinos from reactors},
pdfauthor={Marica Baldoncini}]{hyperref}

\begin{document}

\title{Reference worldwide model for antineutrinos from reactors} 

\author{Marica Baldoncini}
 \affiliation{Dipartimento di Fisica e Scienze della Terra, Università degli Studi di Ferrara,\\ Via Saragat 1-44122, Ferrara, Italy}
  \affiliation{INFN, Sezione di Ferrara, Via Saragat 1-44122, Ferrara, Italy}

\author{Ivan Callegari}
   \affiliation{Laboratori Nazionali di Legnaro, INFN, Viale dell'Università 2-35020 Legnaro, Italy}
   
\author{Giovanni Fiorentini}
  \affiliation{Dipartimento di Fisica e Scienze della Terra, Università degli Studi di Ferrara,\\ Via Saragat 1-44122, Ferrara, Italy}
  \affiliation{INFN, Sezione di Ferrara, Via Saragat 1-44122, Ferrara, Italy}
  \affiliation{Laboratori Nazionali di Legnaro, INFN, Viale dell'Università 2-35020 Legnaro, Italy}

 \author{\\Fabio Mantovani}
   \affiliation{Dipartimento di Fisica e Scienze della Terra, Università degli Studi di Ferrara,\\ Via Saragat 1-44122, Ferrara, Italy}
  \affiliation{INFN, Sezione di Ferrara, Via Saragat 1-44122, Ferrara, Italy}
  
 \author{Barbara Ricci}
   \affiliation{Dipartimento di Fisica e Scienze della Terra, Università degli Studi di Ferrara,\\ Via Saragat 1-44122, Ferrara, Italy}
  \affiliation{INFN, Sezione di Ferrara, Via Saragat 1-44122, Ferrara, Italy}
  
 \author{Virginia Strati}
    \affiliation{Dipartimento di Fisica e Scienze della Terra, Università degli Studi di Ferrara,\\ Via Saragat 1-44122, Ferrara, Italy}
  \affiliation{Laboratori Nazionali di Legnaro, INFN, Viale dell'Università 2-35020 Legnaro, Italy}

 \author{Gerti Xhixha}
  \affiliation{Laboratori Nazionali di Legnaro, INFN, Viale dell'Università 2-35020 Legnaro, Italy}


\begin{abstract}

Antineutrinos produced at nuclear reactors constitute a severe source of background for the detection of geoneutrinos, which bring to 
the Earth's surface information about natural radioactivity in the whole planet.
In this framework we provide a reference worldwide model for antineutrinos from reactors, 
in view of reactors operational records yearly published by the International Atomic Energy Agency (IAEA).
We evaluate the expected signal from commercial reactors for ongoing (KamLAND and Borexino), planned (SNO+) and proposed 
(Juno, RENO-50, LENA and Hanohano) experimental sites.
Uncertainties related to reactor antineutrino production, propagation and detection processes are estimated using a Monte Carlo based 
approach, which provides an overall site dependent uncertainty on the signal in the geoneutrino energy window on the order of 
3\%.

We also implement the off-equilibrium correction to the reference reactor spectra associated with the long-lived isotopes and we estimate a 
2.4\% increase of the unoscillated event rate in the geoneutrino energy window due to the storage of spent nuclear fuels in the cooling 
pools. 
We predict that the research reactors contribute to less than 0.2\% to the commercial reactor signal in the investigated 14
sites. We perform a multitemporal analysis of the expected reactor signal over a time lapse of 10 years using reactor operational records 
collected in a comprehensive database published at \url{www.fe.infn.it/antineutrino}.

\end{abstract}


\maketitle

\section{\label{s:intro}INTRODUCTION}

The existence of antineutrinos was first theorized in 1930 by Pauli, who attempted to explain the continuous electron energy distribution 
in beta decay as due to the emission of a third light, weakly interacting neutral particle. This prediction was confirmed in 
1956 by Reines and Cowan in the Savannah River Experiment, in which Inverse Beta Decay (IBD) reactions caused by electron 
antineutrinos from nuclear reactors were observed for the first time \cite{cowan_detection_1956}. From then on, antineutrinos from nuclear 
reactors have played a crucial role in exploring neutrino physics, with respect to both the standard three-flavor neutrino oscillations and 
possible signatures of non-standard neutrino interactions.

The observation of reactor antineutrino disappearance by the KamLAND (KL) experiment in 2005 \cite{araki_measurement_2005} confirmed the 
neutrino oscillation as the mechanism behind the solar neutrino deficit identified in 2001 by the SNO experiment 
\cite{ahmad_measurement_2001}, opening the way to precise estimates of the oscillation parameters, as the recent determination of the 
non-zero value of $\theta_{13}$.
Moreover, recent results from reactors 
pointed out an apparent 6\% deficit of electron antineutrinos, referred to as the reactor antineutrino anomaly, which could be compatible 
with the existence of a fourth (sterile) neutrino \cite{abazajian_light_2012}. 

Short-baseline and long-baseline reactor experiments, characterized respectively by a reactor-detector distance small/long in comparison 
with a length scale on the order of 1 km, 
provided significant improvements in the accuracy of neutrino oscillation parameters 
\cite{an_improved_2013, ahn_observation_2012, abe_indication_2012, gando_reactor_2013}. 
Thanks to the remarkable progresses in the neutrino field over the last decades, the possibility of applying neutrino detection 
technologies for safeguard purposes is seriously under investigation \cite{SNIF}.
In the neutrino puzzle, the feasibility of reactor 
antineutrino experiments at medium baseline is currently being explored with the intent of probing neutrino oscillation parameters both at 
short and long wavelength and of potentially investigating interference effects related to the mass hierarchy \cite{capozzi_neutrino_2014}.

Concurrently, antineutrinos produced at nuclear reactors constitute a severe source of background for the detection of geoneutrinos, i.e. 
the electron antineutrinos produced in beta minus decays along the $^{238}$U and $^{232}$Th decay chains. 
As the energy spectrum of antineutrinos from nuclear reactors overlaps with the spectrum of geoneutrinos,  a careful analysis of the 
expected reactor signal at specific experimental sites is mandatory to establish the sensitivity to geoneutrinos. 
Geoneutrinos are a real time probe of the Earth's interior as their flux at the terrestrial surface depends on the amount and on the 
distribution of $^{238}$U and $^{232}$Th naturally present in the crust and in the mantle, which are thought to be the main reservoirs of 
these radioisotopes \cite{fiorentini_geo-neutrinos_2007}. 
The first experimental evidence of geoneutrinos dates from 2005, when the KL Collaboration claimed the observation of four 
events associated with $^{238}$U and five with $^{232}$Th decay chains \cite{araki_experimental_2005}. 
Recent results from the KL and Borexino (BX) experiments provided quantitative measurements of the geoneutrino signal 
(116$^{+28}_{-27}$ observed events in a total live-time of 2991 days for KL \cite{gando_reactor_2013} and from 14.3 $\pm$ 4.4 geoneutrino 
events in 1353 days for BX \cite{borexino_collaboration_measurement_2013}), important for discriminating among different Earth 
compositional models.

The crustal contribution to the geoneutrino signal can be inferred from direct geochemical and geophysical surveys, while the mantle 
contribution is totally model-dependent. A better discrimination among different compositional models of the Bulk Silicate Earth (BSE), 
referred to as ``cosmochemical'', ``geochemical'', and ``geodynamical'' \cite{sramek_geophysical_2013}, can be attained by combining the 
results from several sites \cite{huang_reference_2013}. Therefore, new measurements of geoneutrino fluxes are highly awaited from 
experiments entering operation, such as SNO+ \cite{maneira_sno+_2013}, or proposed to the scientific community, such as Juno 
\cite{li_overview_2014}, RENO-50 \cite{proposal_reno_50}, LENA \cite{wurm_next-generation_2012}, Hanohano \cite{cicenas_hanohano_2012}, 
Homestake \cite{tolich_geoneutrino_2006} and Baksan \cite{domogatsky_neutrino_2005}.

Electron antineutrinos are currently detected in liquid scintillation detectors via the 
IBD reaction on free protons
\begin{equation}
\bar{\nu}_{e}+p\rightarrow n+e^{+}
\end{equation}
which has an energy threshold of 1.806 MeV.
As the antineutrino detection depends on several experimental parameters (e.g. the fiducial volume), expressing both geoneutrino 
and reactor antineutrino signals in terms of detector independent quantities allows the comparison of signals measured at 
different experiments and originating from different sources. Therefore, event rates are quoted in Terrestrial Neutrino Units (TNU) 
\cite{fiorentini_geo-neutrinos_2007}, corresponding to one event per 10$^{32}$ target protons per year, which are practical units 
as liquid scintillator mass is on the order of one kton ($\sim 10^{32}$ free protons) and the exposure times are typically on the order of 
a few years.

\begin{figure}
\includegraphics[width=\linewidth]{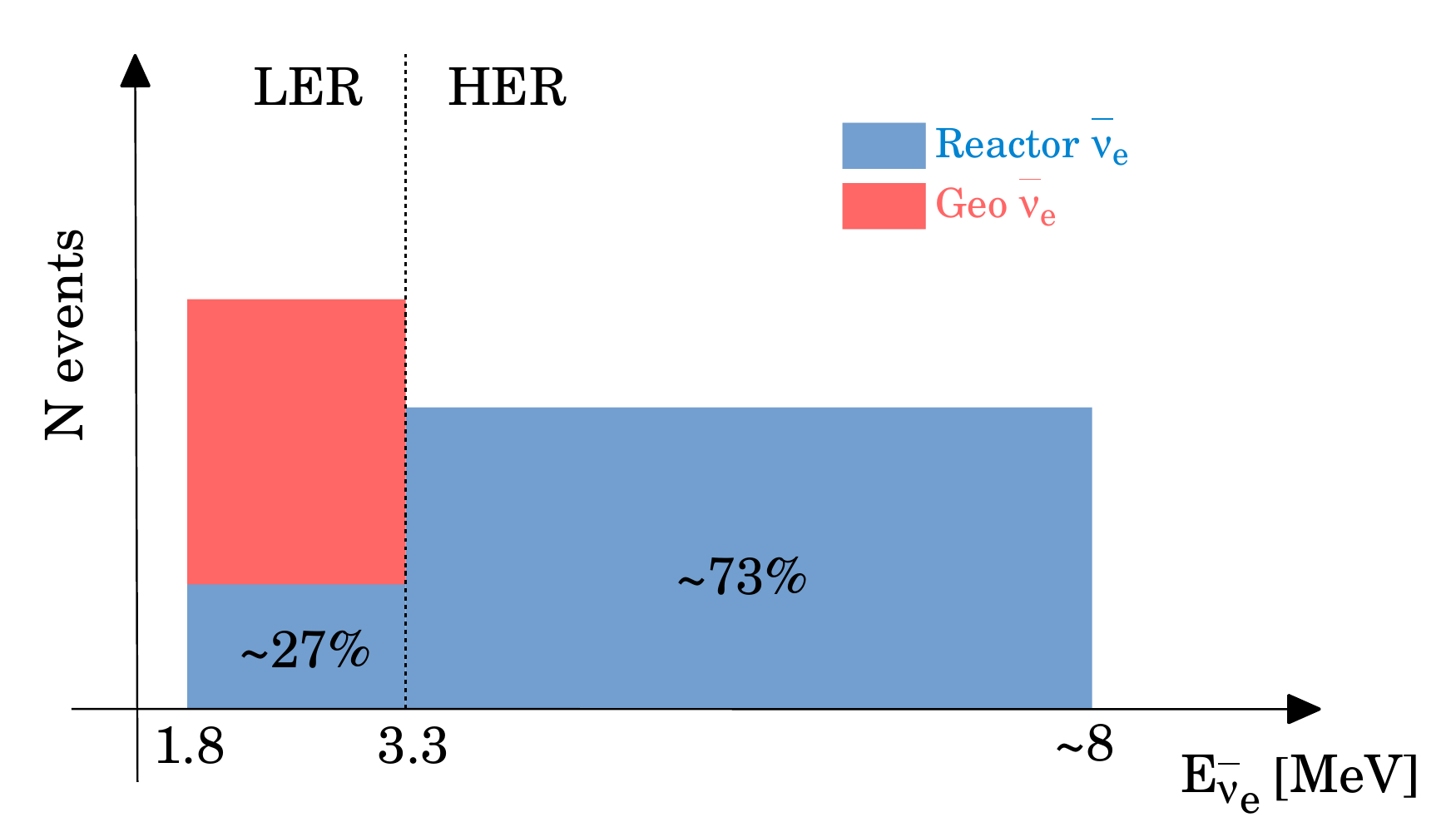}

\caption{\label{fig_blu_rosa}A sketch of the expected reactor signal in the Low Energy Region (LER) and in the High Energy Region (HER). The 
reactor signal in the HER is crucial for modeling the reactor contribution in the LER, and 
therefore for extracting information on geoneutrinos. The reactor contribution to the signal changes according to the different reactor 
operational conditions, while the geoneutrino component is time independent.}
\end{figure}

Considering that the reactor antineutrino spectrum extends beyond the endpoint of that of the geoneutrinos, we observe a significant 
overlap between geoneutrino and reactor signals in the geoneutrino energy window (\figname{\ref{fig_blu_rosa}}), where generally about 27\% 
of the total reactor events are registered.
The boundaries of this energy range, also specified as Low Energy Region (LER), are defined by the detection reaction threshold and by the 
maximum energy of emitted geoneutrinos, occurring in the $^{214}$Bi beta minus decay (3.272 MeV) \cite{fiorentini_nuclear_2010}. The 
High Energy Region (HER) extends from the upper edge of the LER to the endpoint of the reactor antineutrino spectrum. In this framework, 
modeling the predicted signal in the HER where only reactor events are expected, is of decisive importance for understanding the reactor 
contribution in the LER. In particular, the ratio R$_{\text{LER}}$/G between the predicted reactor signal in the LER (R$_{\text{LER}}$) and 
the expected geoneutrino signal (G) can be considered as a figure of merit for assessing the discrimination power on geoneutrinos at a 
specific location. 

The focus of this paper is the calculation of the antineutrino signal from nuclear power plants, as fundamental background for geoneutrino 
measurements. Our work is structured as follows. In \secname{\ref{ingredients}} we present all the required ingredients for the calculation 
of the expected reactor signal, where we distinguish the three distinctive antineutrino life stages, i.e. production, propagation and 
detection. In \secname{\ref{sec:uncertainties}} we describe the Monte Carlo based approach we adopt to estimate the global uncertainty on 
the reactor signal, together with the relative contributions related to each input quantity of the calculation. In \secname{\ref{results}} 
we present 10 years (2003 – 2013) of reactor antineutrino signals at KL and BX, along with updated estimates of the expected reactor signals 
throughout the world, with a particular focus on ongoing and proposed experimental sites. In \secname{\ref{conclusions}} we summarize the 
main results of our work.

\section{\label{ingredients}Ingredients in the calculation of the reactor antineutrino signal}

The dominating background in geoneutrino studies is due to electron antineutrinos produced at nuclear power plants, which are the strongest 
man-made antineutrino sources. With an average energy released per fission of approximately 200 MeV and 6 antineutrinos 
produced along the beta minus decay chains of the neutron-rich unstable fission products, 6 $\cdot$ 10$^{20}$ $\bar{\nu}$/s are emitted 
from a reactor having a thermal power of 3 GW. 
Evaluating the reactor antineutrino signal at a given location requires the knowledge of several ingredients, necessary for modeling the 
three reactor antineutrino life stages: production at reactor cores, propagation to the detector site and detection in liquid scintillation 
detectors via the IBD reaction. 

In our calculation, we consider all not movable operational reactors in the world used for commercial and research purposes. 
Hundreds of naval nuclear reactors with thermal power on the order of some hundreds of MW drive submarines, aircraft carriers and 
icebreakers cruising around the world \cite{Hirdaris_2014}. A discussion of the potential effect due to nuclear propelled vessels on 
neutrino measurements is provided in \cite{gratta}. 

A comprehensive database dating back to 2003 has been compiled that contains the main features of each operational reactor core. 
The database is available at \url{www.fe.infn.it/antineutrino} and we plan to update it every year. 
The database structure is described in the Appendix.

\subsection{\label{spectra}Spectra of antineutrinos produced at reactor cores}

\begin{table*}
\caption{\label{tab:fission_fraction}$^{235}$U, $^{238}$U, $^{239}$Pu, and $^{241}$Pu fission/power fractions for PWRs, BWRs, 
GCRs, 
LWGRs, PHWRs and for reactors burning MOX, taken from literature references.}
\begin{ruledtabular}
\renewcommand{\arraystretch}{1.15}
\begin{tabular}{ >{\centering\arraybackslash} m {0.9in}  >{\centering\arraybackslash} m{0.7in}  >{\centering\arraybackslash} 
m{0.8in} >{\centering\arraybackslash} m{0.8in}  >{\centering\arraybackslash} m{0.8in}  >{\centering\arraybackslash} m{0.8in}  
>{\centering\arraybackslash} m{1.7in}}
Reactor classes &Fractions& $^{235}$U& $^{238}$U&$^{239}$Pu&$^{241}$Pu&Reference \bigstrut \\
\hline
\multirow{22}{\linewidth}{\parbox{\linewidth}{PWR\\BWR\\LWGR\\GCR}}& 
\multicolumn{1}{c}{\multirow{21}{0.7in}{\parbox{0.7in}{$f_{i}$}}}&0.538&0.078&0.328&0.056& 
\multirow{10}{1.7in}{\parbox{1.7in}{\citet{mention_reactor_2011}}} \bigstrut[t]\\
& &0.614&0.074&0.274&0.038&\\
& &0.620&0.074&0.274&0.042&\\
& &0.584&0.068&0.298&0.050&\\
& &0.543&0.070&0.329&0.058&\\
& &0.607&0.074&0.277&0.042&\\
& &0.603&0.076&0.276&0.045&\\
& &0.606&0.074&0.277&0.043&\\
& &0.557&0.076&0.313&0.054&\\
& &0.606&0.074&0.274&0.046&\\
\cline{3-7}
& &0.488&0.087&0.359&0.067&\citet{abe_indication_2012}\bigstrut \\
\cline{3-7}
& &0.580&0.074&0.292&0.054&\multirow{3}{1.7in}{\parbox{1.7in}{\citet{djurcic_uncertainties_2009}}} \bigstrut[t]\\
& &0.544&0.075&0.318&0.063&\\
& &0.577&0.074&0.292&0.057&\\
\cline{3-7}
& &0.590&0.070&0.290&0.050&\citet{kopeikin_reactor_2004}\bigstrut \\
\cline{3-7}
& &0.570&0.078&0.295&0.057&\citet{abe_precision_2008}\bigstrut \\
\cline{3-7}
& &0.568&0.078&0.297&0.057&\citet{eguchi_first_2003}\bigstrut \\
\cline{3-7}
& &0.563&0.079&0.301&0.057&\citet{araki_measurement_2005}\bigstrut\\
\cline{3-7}
& &0.650&0.070&0.240&0.040&\multirow{3}{1.7in}{\parbox{1.7in}{\citet{kopeikin_flux_2012}}} \bigstrut[t]\\
& &0.560&0.070&0.310&0.060&\\
& &0.480&0.070&0.370&0.080&\\
\cline{2-7}
& $p_{i}$&0.560&0.080&0.300&0.060&\citet{bellini_observation_2010}\bigstrut\\
\hline
MOX& $p_{i}$&0.000&0.081&0.708&0.212&\citet{bellini_observation_2010}\bigstrut\\
\hline
PHWR&$p_{i}$&0.543&0.411&0.022&0.024&\citet{borexino_collaboration_measurement_2013}\bigstrut[t]\\
\end{tabular}
\end{ruledtabular}
\end{table*}

The operating principle of nuclear power reactors lies in the generation of heat by the neutron-induced fissions of U and Pu isotopes and by 
the subsequent decays of unstable fission fragments. In a typical reactor, more than 99.9\% of antineutrinos above the IBD energy threshold 
are emitted in large Q-value beta decays of unstable daughter fragments that originated in the fission process of just four isotopes: 
$^{235}$U, $^{238}$U, $^{239}$Pu and $^{241}$Pu \citep{djurcic_uncertainties_2009}. Therefore, the antineutrino spectrum produced by a given 
reactor can be expressed, in units of $\bar{\nu}$/MeV/fission, as:
\begin{equation}
 \Lambda(E_{\bar{\nu}})=\sum_{i=1}^{4}f_{i}\lambda_{i}(E_{\bar{\nu}})
\end{equation}
where $\lambda_{i}(E_{\bar{\nu}})$ and $f_{i}$ are respectively the antineutrino emission spectrum normalized to 
one fission process and the fission fraction for the $i$-$th$ isotope, where $i =^{235}$U, $^{238}$U, $^{239}$Pu and $^{241}$Pu. 
In literature, the different fuel isotope contributions to the generated thermal power are expressed as fission fractions or as power 
fractions, which have to be considered as different physical quantities. The fission fraction $f_{i}$ is defined as a relative fission 
yield, i.e., as the fraction of fissions produced by the $i$-$th$ isotope. This quantity is related to the reactor thermal power by the 
following energy relation:
\begin{align} 
P_{th}=R\langle Q\rangle=&R\sum_{i=1}^{4}f_{i}Q_{i}
\end{align}
where $R$ is total fission rate (number of fissions per unit time) and $\langle Q\rangle$ is the average energy released per fission. 
The same energy relation can be expressed in terms of the power fractions $p_{i}$, corresponding to the fraction of the total 
thermal power produced by the fission of the $i$-$th$ isotope:
\begin{align} 
P_{th}=p_{i}P_{i}=&p_{i}Rf_{i}Q_{i}
\end{align}
where $P_{i}$ is the thermal power generated by isotope $i$. Accordingly, the following relation between power fractions and fission 
fractions 
holds:
\begin{equation} 
p_{i}=\frac{f_{i}Q_{i}}{\sum\limits_{i=1}^{4}f_{i}Q_{i}}
\end{equation}

During the power cycle of a nuclear reactor, the composition of the fuel changes as Pu isotopes are bred and U is consumed: thus, the 
power (fission) fractions are time-dependent quantities. Fuel isotope contributions also depend on the burn-up technology adopted in the 
given reactor core as different core types are characterized by different fuel compositions, which in turn give rise to different 
isotope contributions to the total thermal power. 

The nuclear reactor operation relies on the use of cooling and moderating materials, which should be as safe and as cheap as possible. 
Typical coolants include materials such as water or gas which, due to their high thermal capacity, allow the collection and transfer of the 
energy released in the fission processes, while moderators are exploited to slow down the neutrons resulting from the fission processes to 
thermal energies in order to maintain the fission chain. Ordinary water is the most common moderator material: indeed, since hydrogen has a 
mass almost identical to that of the incident neutron, a single neutron-hydrogen collision can reduce the speed of the neutron 
substantially. However, due to the relatively high neutron capture cross section, reactors using light water as moderator (such as 
Pressurized Water Reactors (PWRs) and Boiling Water Reactors (BWRs)) require the adoption of enriched uranium as nuclear fuel, with a 
typical enrichment level of $^{235}$U ranging from 2\% to 
5\% \cite{bemporad_reactor-based_2002}. Pressurized Heavy Water Reactors (PHWRs) use heavy water as both moderator and 
coolant: due to the smaller neutron capture cross section with respect to ordinary water, PHWRs can burn natural uranium. However, as the 
reactor design is flexible and allows the use of advanced fuel cycles, using slightly enriched uranium, recovered
uranium, Mixed OXide fuel (MOX), thorium fuels, and others \cite{rouben_candu} is possible. Gas Cooled Reactors (GCRs)\footnote{Modern 
reactors using gas as cooling material and graphite moderated are also referred to as AGRs (Advanced Gas-cooled Reactors).} 
and Light Water Graphite Reactors (LWGRs) exploit graphite as moderator, which allows the adoption of lower uranium enrichment levels, 
typically between 2.2\% and 2.7\% \cite{podvig_history_2011,nonbol_agr}.
Few tens of reactors (mainly located in Europe) use MOX, which is a mix of more than one 
oxide of fissile material and usually consists of plutonium recovered from spent nuclear fuel, blended with natural uranium, reprocessed 
uranium or depleted uranium. Generally, approximately 30\% of the total power of these reactors comes from the MOX fuel, while the 
remaining 70\% of the power is produced by standard fuel \cite{bellini_observation_2010}.

In our calculation of the emitted reactor antineutrino spectrum $\Lambda(E_{\bar{\nu}})$ we distinguish reactor classes according to the 
employed nuclear fuel. In \tabname{\ref{tab:fission_fraction}} we report typical fission/power fractions, together with 
the corresponding literature reference. PHWRs power fractions refer to reactors burning natural 
uranium \cite{canada_1997}; therefore, we assign PWRs, BWRs, LWGRs and GCRs to the same class of enriched uranium 
burning reactors.

\begin{table}
\caption{\label{Qi}Energy released per fission $Q_{i}$ for $^{235}$U, $^{238}$U, $^{239}$Pu, and $^{241}$Pu taken from 
\citet{ma_improved_2013}.}
\begin{ruledtabular}
\renewcommand{\arraystretch}{1.15}
\begin{tabular}{c c}
Fissile isotope & $Q_{i}$(MeV) \bigstrut \\
\colrule
$^{235}$U & 202.36 $\pm $ 0.26 \bigstrut [t]\\
$^{238}$U  & 205.99 $\pm $ 0.52\\
$^{239}$Pu  & 211.12 $\pm $ 0.34\\
$^{241}$Pu &214.26 $\pm$ 0.33
\end{tabular}
\end{ruledtabular}
\end{table}

\begin{table*}
\caption{\label{tab:mueller}Coefficients of the polynomial of order 5 used as argument of the exponential function for the analytical 
expression of the 
antineutrino spectra for $^{235}$U, $^{238}$U, $^{239}$Pu, and $^{241}$Pu, taken from \citet{mueller_improved_2011}.}
\begin{ruledtabular}
\renewcommand{\arraystretch}{1.2}
\begin{tabular}{c c c c c c c}
Fissile isotope & $a_{1}$  & $a_{2}$ & $a_{3}$ & $a_{4}$ & $a_{5}$ & $a_{6}$ \bigstrut \\
\hline
$^{235}$U & 3.217&-3.111&1.395&-3.690$(10^{-1})$&4.445(10$^{-2}$)&-2.053(10$^{-3}$) \bigstrut[t] \\
$^{238}$U  & 4.833(10$^{-1}$)&1.927(10$^{-1}$)&-1.283(10$^{-1}$)&-6.762(10$^{-3}$)&2.233(10$^{-3}$)&-1.536(10$^{-4}$)\\
$^{239}$Pu  &6.413&-7.432&3.535&-8.820(10$^{-1}$)&1.025(10$^{-1}$)&-4.550(10$^{-3}$) \\
$^{241}$Pu & 3.251&-3.204&1.428&-3.675(10$^{-1}$)&4.254(10$^{-2}$)&-1.896(10$^{-3}$)\\
\end{tabular}
\end{ruledtabular}
\end{table*}

The contribution to the reactor thermal power given by each fuel isotope depends on its specific fission fraction as well as on the energy 
released per fission $Q_{i}$, which is obtained by: 
\begin{equation}\label{eq:Qi}
 Q_{i}=E_{tot}^{\text{ }i}-\langle E_{\bar{\nu}}\rangle^{i}-\Delta E_{\beta\gamma}^{\text{ }i}+E_{nc}^{\text{ }i}
\end{equation}
where $E_{tot}^{\text{ }i}$ is the total energy produced in a fission process, starting from the moment the neutron that induces the process 
is absorbed until all of the unstable fission fragments have undergone beta decays; $\langle E_{\bar{\nu}}\rangle^{i}$ is the mean energy 
carried away by antineutrinos produced in the beta decays of fission fragments; $\Delta E_{\beta\gamma}^{\text{ }i}$ is the energy of beta 
electrons and photons that, on average, does not contribute to the reactor energy during the operation of the core; $E_{nc}^{\text{ }i}$ is 
the energy released in neutron capture (without fission) by the reactor core materials \cite{ma_improved_2013}. 
In \tabname{\ref{Qi}} we list the energies released per fission adopted in the calculation of the reactor antineutrino spectrum, which have 
been computed by \citet{ma_improved_2013} following the approach described in \equatname{\ref{eq:Qi}}.

The distribution of the fission products of uranium or plutonium involves hundreds of nuclei, each of them contributing to 
$\lambda_{i}(E_{\bar{\nu}})$ through various beta decay chains. Thus, the total antineutrino spectrum is the result of the sum 
of thousands of beta branches, weighted by the branching ratio of each transition and by the fission yield of the parent nucleus. The two 
traditional ways for predicting the total antineutrino spectrum are the summation and the conversion methods. The  summation procedure 
reconstructs the beta spectra using available nuclear databases as the sum of the branch-level beta spectra of all the daughter isotopes 
and 
then converts the beta spectra in antineutrino spectra. The conversion technique relies on direct measurements of the beta 
spectra and exploits the energy conservation law between the two leptons involved in the beta minus decay: 
\begin{equation}
 E_{e}+E_{\bar{\nu}}=E_{0}
\end{equation}
where $E_{0}$ is the endpoint of the beta transition.

In the 1980s, measurements of the total beta spectra of fissile isotopes were performed at the Laue-Langevin Institute (ILL) in 
Grenoble where thin 
target foils of $^{235}$U, $^{239}$Pu and $^{241}$Pu were exposed to an intense thermal neutron flux and the beta spectra of the unstable 
fragments were measured \cite{schreckenbach_determination_1985,von_feilitzsch_experimental_1982, hahn_antineutrino_1989}. 
These spectra act as benchmarks for the summation calculations and are direct inputs for the conversion method.
As $^{238}$U undergoes fission when bombarded by fast neutrons, its beta spectrum 
could not be measured in the thermal flux of ILL. Recently, an experiment was performed at the neutron source FRM II in 
Garching to determine the cumulative antineutrino spectrum of the fission products of  $^{238}$U \cite{haag_experimental_2014}. 

In this work, we adopt as reference model the one published by \citet{mueller_improved_2011}, where the spectra of all four contributing 
isotopes 
are consistently given in terms of the exponential of a polynomial of order 5, as stated in 
\equatname{\ref{eq:mueller}}. \citet{mueller_improved_2011} derive the $^{235}$U, $^{239}$Pu and $^{241}$Pu spectra based on a 
mixed approach that combines the accurate reference of the ILL electron spectra with the physical distribution of beta branches provided 
by the 
nuclear databases, and calculates the $^{238}$U spectrum via a pure summation method.
\begin{equation}\label{eq:mueller}
 \lambda_{i}(E_{\bar{\nu}})=\text{exp} \left(\sum_{p=1}^{6}a_{p}^{i}E_{\bar{\nu}}^{p-1}\right) 
\end{equation}
In \tabname{\ref{tab:mueller}}, the coefficients of the polynomial function used in the parametrization of the reactor antineutrino 
spectrum 
generated by each fuel isotope are listed.

A reactor operational time profile is a required input for estimating the number of fissions occurring in a given time interval. The Load 
Factor ($LF$) is the percentage quantity expressing the effective working condition of a core in a specific period of the operating cycle 
and is defined as the ratio
\begin{equation}
 LF=100\cdot\frac{EG}{REG}
\end{equation}
where $EG$ is the net electrical energy produced during the reference period as measured at the unit outlet terminals, i.e. after 
subtracting the electrical energy taken by auxiliary units, while $REG$ is the net electrical energy that would have been supplied to 
the grid if the unit were operated continuously at the reference power unit during the whole reference period 
\cite{iaea_operating_experience_2013}. 
Load factor data are published by the IAEA \cite{iaea_operating_experience_2013}, both on a monthly timeline and as an annual average. In 
our calculation we assume that published values of electrical load factors are equal to thermal load factors.

The spectrum of reactor antineutrinos emitted by a reactor core having a thermal power $P_{th}$  and operating with a load factor $LF$ can 
thus be 
evaluated according to \equatname{\ref{eq:total_spectrum}}.
\begin{equation}\label{eq:total_spectrum}
 S(E_{\bar{\nu}})=P_{th}LF \sum_{i=1}^{4} \frac{p_{i}}{Q_{i}}\lambda_{i}(E_{\bar{\nu}})
\end{equation}

\subsection{\label{propagation} Propagation of antineutrinos during their movement to detector}

The demonstration of the separate identity of muon and electron neutrinos \cite{danby_observation_1962}, the discovery 
of the tauonic neutrino \cite{kodama_observation_2001} and the measurement of the decay width of the Z boson at LEP \cite{groom_2000} 
endorsed 
the Standard Electroweak Model (SEM) as the most reasonable theory describing neutrino physics, according to which neutrinos exist in three 
light (with masses smaller than 1/2 M$_{\text{Z}}$) flavors and separate lepton numbers for electron, muon, and tau flavors are conserved.
Nevertheless, an observed deficit in the solar neutrino flux with respect to the prediction of the Standard Solar Model started questioning 
the SEM, until the neutrino flavor change was definitely identified in 2001 by the SNO experiment \cite{ahmad_measurement_2001} and 
subsequently the KL experiment provided clear evidence of the neutrino oscillatory nature \cite{gando_2011}.

At present, most experimental results on neutrino flavor oscillation agree with a three neutrino scenario, where weak neutrino 
eigenstates, i.e. flavor eigenstates $(\nu_{e}, \nu_{\mu}, \nu_{\tau})$ mix with the mass eigenstates $(\nu_{1}, \nu_{2}, \nu_{3})$ 
via three mixing angles $(\theta_{12}, \theta_{13} , \theta_{23})$ and a possible CP-violating phase $\delta$. 
Therefore, to establish the reactor antineutrino flux at a given site, it 
is necessary to consider the survival probability of the electron antineutrino, which can be expressed (assuming that antineutrinos 
propagate in vacuum) in terms of the mass-mixing 
oscillation parameters $(\delta m^{2} , \theta_{12} , \theta_{13})$ as stated in \citet{fiorentini_mantle_2012}:
\begin{align}\label{eq:pee}
P_{ee}(E_{\bar{\nu}}, L) = & cos^{4}(\theta_{13})\left(1-sin^{2}(2\theta_{12})sin^{2}\left(\frac{\delta m^{2} 
L}{4E_{\bar{\nu}}}\right)\right) \nonumber \\& + sin^{4}(\theta_{13}),
\end{align}
where $L$ and $E_{\bar{\nu}}$ are the antineutrino path length and energy in natural units\footnote{The 3 flavor vacuum survival 
probability in principle depends on the difference between the squared masses $\Delta m^{2} = m^{2}_{3} − (m^{2}_{1} +m^{2}_{2})/2$, 
according to a relationship that is not invariant under a change of hierarchy (where $\Delta m^{2} > 0$ and $\Delta m^{2} <0$ correspond 
respectively to the normal and inverted hierarchy scenarios). In any case, the $\Delta m^{2}$ dependence of the survival probability is 
negligible for $L>> 50$ km \cite{capozzi_neutrino_2014}. Considering the quality of the inputs used for our calculation, the differences on 
the expected signal due to the use of two survival probabilities ($\Delta m^{2}$ dependent and $\Delta m^{2}$ not dependent) are 
negligible, also in the case of JUNO and RENO-50. The most general survival probability should be used for a spectral shape analysis, but it 
goes beyond the scope of this paper.}.  

In our calculation we adopt the updated values on neutrino oscillation parameters, obtained by \citet{capozzi_status_2014} from a global 
fit to data provided by different experiments.\footnote{During the refereeing process of the present work, new 
releases of oscillation parameters appeared in \citet{forero_2014} and \citet{gonzalez_2014}, affecting mainly the central values and the 
uncertanties on sin$^{2}(\theta_{12})$ and sin$^{2}(\theta_{13})$. A check on the expected reactor signals shows central values variations 
within 1$\sigma$ reported in Table \ref{tab:15_sites}, together with a $\sim$20\% decrease on the associated uncertainties. In this 
perspective, our estimations in Table \ref{tab:uncertainties} and Table 
\ref{tab:15_sites} can be considered conservative.}
The data combined analysis provides N$\sigma$ curves of the 3$\nu$ oscillation 
parameters, whose degree of linearity and symmetry is strictly related to the Gaussian nature of the probability distribution associated 
with that parameter. On the basis of Fig. 3 of \citet{capozzi_status_2014}, we assume $(\delta m^{2} , \theta_{12} , \theta_{13})$ as 
described by Gaussian Probability Density Functions (PDF) and we adopt as central values and 1$\sigma$ uncertainties the values reported in 
\tabname{\ref{tab:oscillation_param}}, where, conservatively, the 1$\sigma$ value has been selected as the maximum between $\sigma^{+}$ and 
$\sigma^{-}$ for each parameter distribution.  
\begin{table}
\caption{\label{tab:oscillation_param}The 3$\nu$ mass-mixing parameters entering the electron antineutrino
survival probability equation, adapted from \citet{capozzi_status_2014}}.

\begin{ruledtabular}
\renewcommand{\arraystretch}{1.15}
\begin{tabular}{c c}
Oscillation parameter & Central value $\pm 1\sigma$ range \bigstrut \\
\hline
$\delta m^{2}$ (eV$^{2}$)& 7.54 $\pm $0.26 $(10^{-5})$ \bigstrut[t] \\
sin$^{2}(\theta_{12})$ & 3.08 $\pm $0.17 $(10^{-1})$\\
sin$^{2}(\theta_{13})$  & 2.34 $\pm $0.20 $(10^{-2})$\\
\end{tabular}
\end{ruledtabular}
\end{table}

We investigated the matter effect concerning the antineutrino propagation from the reactor to 
the experimental site by adopting the Earth density profile as published in \citet{Dziewonski:1981xy}.
The matter effect on the signal varies according to the investigated experimental site, giving a maximum contribution of 0.7\% at Hawaii. 
In any case, it can be considered negligible at 
1$\sigma$ level with respect to the overall uncertainties reported in \tabname{\ref{tab:15_sites}}. 

With respect to the antineutrino pathlength, we evaluate $L$ as the distance from the reactor to the experimental site using an ellipsoid 
as 
geometrical shape of the Earth. We use a= 6378136.6 m and b= 6356751.8 m as equatorial radius and polar radius, 
respectively \cite{iers_2003}. 

\subsection{\label{detection}Detection of antineutrinos}

The components presented in the last two sections allow the modeling of the expected (oscillated) reactor antineutrino flux at a given 
experimental site. To determine the predicted signal, it is necessary to account for the detection process via the IBD reaction on 
free protons. The IBD reaction effectiveness in antineutrino detection is the result of the relatively large reaction cross section 
(on the order 
of 10$^{-42}$cm$^{2}$), the feasibility of building large detectors (as materials rich in free protons, such as water and 
hydrocarbons, are relatively cheap) and the possibility of reducing backgrounds, which is possible due to the correlation between 
the prompt positron 
annihilation signal and the delayed neutron capture signal \cite{jocher_theoretical_2013}. 
In this work, we use for the parametrization of the IBD reaction cross section 
the expression given by \citet{strumia_precise_2003}:

\begin{widetext}
\begin{align}\label{eq:sigma_IBD}
 \sigma_{IBD}(E_{\bar{\nu}})& 
=10^{-43}cm^{2}p_{e}E_{e}E_{\bar{\nu}}^{-0.07056+0.02018lnE_{\bar{\nu}}-0.001953ln^{3}E_{\bar{\nu}}}\text{,}& 
E_{e}&=E_{\bar{\nu}}-\Delta\text{,}&p_{e}&=\sqrt{E_{e}^{2}-m_{e}^{2}}\text{,}
 \end{align}
\end{widetext}
where $E_{e}$ is the positron energy, $\Delta= m_{n}-m_{p} \approx 1.293$ MeV, $p_{e}$ is the positron momentum, $m_{e}= 0.511$ MeV is 
the positron mass. The final equation for the evaluation of the antineutrino signal from reactors is obtained considering the contribution 
at a given 
experimental site given by all operating reactors in the world, as stated in \equatname{\ref{eq:signal}}

\begin{widetext}
\begin{equation}\label{eq:signal}
 N_{tot}=\varepsilon N_{p} \tau \sum_{i=1}^{N_{reactor}}\frac{P_{th}^{i}}{4 \pi L_{i}^{2}}\langle LF_{i}\rangle\int dE_{\bar{\nu}} 
\sum_{k=1}^{4}\frac{p_{k}}{Q_{k}} \lambda_{k}(E_{\bar{\nu}})P_{ee}(E_{\bar{\nu}},L_{i})\sigma_{IBD}(E_{\bar{\nu}})
\end{equation}
\end{widetext}
where $\varepsilon$ is the detector efficiency, $N_{p}$ is the number of free target protons, $\tau$ is the exposure time, $\langle 
LF_{i}\rangle$ is the average load factor of the $i$-$th$ reactor over the given exposure time and $L_{i}$ is the reactor-detector 
distance. 
We evaluate the reactor antineutrino signal in TNU and therefore assume a total number of free protons equal to $N_{p} = 10^{32}$, an 
acquisition time $\tau$ = 3.15 10$^{7}$ s (1 year) and a detector 
efficiency $\varepsilon$ = 1.

\section{\label{sec:uncertainties}Estimation of uncertainties}

The calculation of the reactor antineutrino signal at a given site requires the knowledge of many factors related to reactor physics, in 
terms of reactor operations and of nuclear physics describing the fission process, and to antineutrino physics, which involves both the 
oscillation and the detection mechanisms. Uncertainties with respect to input data contribute with different weights and in different 
ways to 
the uncertainty on the reactor signal. Thus, given the complexity of the model, we used a Monte Carlo based approach to estimate 
the global uncertainty on the reactor signal, together with the relative contributions associated with each component of the calculation.

According to \cite{jcgm:2008:PDMC}, for the evaluation of the uncertainty on the signal due to a specific input 
quantity $X_{i}$ we fix all the components to their central values and conduct a Monte Carlo sampling of $X_{i}$ pseudo random 
values according to their PDFs. With respect to the fission fractions, we assume as central values for the reactor class involving PWRs, 
BWRs, LWGRs and GCRs the set reported in \citet{bellini_observation_2010}.
In \tabname{\ref{tab:uncertainties}}, we summarize the PDFs and the 
associated standard errors for the input quantities included in the propagation of the uncertainties, together with the 
reference from which each parameter has been extracted.
Althought moderate correlations among some signal input quantities (e.g. thermal power and fission fractions) have been investigated by 
\citet{djurcic_uncertainties_2009}, the analysis of their effects is out of the goal of this study as it would require punctual knowledge 
of input data (e.g. stage of burn up of the fuel, effective thermal power). In this framework we treat each parameter as uncorrelated 
with other input quantities. 

The signal uncertainties associated with each single input for 
the KL, BX and SNO+ experiments (see \tabname{\ref{tab:uncertainties}}) are obtained performing 10$^{4}$ calculations of the 
global signal produced by all operating reactors in the world in 2013 and using the reactor antineutrino spectrum provided by 
\citet{mueller_improved_2011}.

With respect to the antineutrino oscillation parameters and the energy released per fission, the same $X_{i}$ sampled value is used for all 
operating reactors for a given global signal calculation. 

The fission fractions are extracted for the single cores for each of the 10$^{4}$ total reactor signal calculations at a given 
experimental site. 
The random sampling of the fission fractions allows to take into account the lack of knowledge concerning the detailed 
fuel composition 
of each reactor as well as the unknown stage of burn-up. The sampling is performed for PWRs, BWRs and GCRs and for the 70\% contribution 
given by standard fuels for reactors using the MOX technology. This is carried out by extracting with equal 
probability one of the 22 sets of 
fission fractions listed in \tabname{\ref{tab:fission_fraction}} (constant Probability Mass Function (PMF)). For 
PHWRs and for the 30\% MOX component the fixed values adopted are those presented in \citet{bellini_observation_2010} 
and \citet{borexino_collaboration_measurement_2013}, listed in \tabname{\ref{tab:fission_fraction}}. 
\begin{table*}
\caption{\label{tab:SNF}LLIs, responsible of the off-equilibrium contribution to the reactor antineutrino spectrum during the reactor 
operating period, together with the SNFs (in the last three rows), which contribute also after the shut down of the reactor. 
$\tau_{1/2}^{P}$, $\tau_{1/2}^{D}$, $E_{\bar{\nu}}^{max\text{}P}$ and $E_{\bar{\nu}}^{max\text{}D}$ are the half-lives and the maximum 
energy of the emitted antineutrino of the parent (P) and daughter (D) nucleus, respectively \cite{nuclear_database}.
$Y_{235}$ and $Y_{239}$ are respectively the daughter cumulative specific yields in percentage per fission event of $^{235}$U 
and $^{239}$Pu, except for the case of $^{93}$Y and $^{97}$Zr which refer to the parent nuclides \cite{kopeikin_flux_2012}.}
\begin{ruledtabular}
\renewcommand{\arraystretch}{1.2}
\begin{tabular}{c c c c c c c c}
P & $\tau_{1/2}^{P}$&$E_{\bar{\nu}}^{max\text{ }P}$ [MeV]&D&$\tau_{1/2}^{D}$&$E_{\bar{\nu}}^{max\text{ }D}$ [MeV]&$Y_{235}$(\%) & 
$Y_{239}$(\%) \bigstrut \\
\colrule
$^{93}$Y&10.18 h&2.895&$^{93}$Zr&1.61 $\cdot 10^{6}$yr&0.091&6.35&3.79 \bigstrut [t]\\
$^{97}$Zr&16.75 h&1.916&$^{97}$Nb&72.1 m&1.277&5.92&5.27\\
$^{112}$Pd&21.03 h&0.27&$^{112}$Ag&3.13 h&3.956&0.013&0.13\\
$^{131m}$Te&33.25 h&/&$^{131}$Te&25.0 m&2.085&0.09&0.20\\
$^{132}$Te&3.204 d&0.24&$^{132}$I&2.295 h&2.141&4.31&5.39\\
$^{140}$Ba&12.753 d&1.02&$^{140}$La&1.679 d&3.762&6.22&5.36\\
\hline
$^{144}$Ce&284.9 d&0.319&$^{144}$Pr&17.28 m&2.998&4.58&3.11 \bigstrut [t]\\
$^{106}$Ru&371.8 d&0.039&$^{106}$Rh&30.07 s&3.541&0.30&3.24 \\
$^{90}$Sr&28.79 yr&0.546&$^{90}$Y&64.0 h&2.280&0.27&0.10 \\

\end{tabular}
\end{ruledtabular}
\end{table*}
Although individual measurements of 
reactor thermal power can reach a sub-percent level accuracy \cite{djurcic_uncertainties_2009,bemporad_reactor-based_2002}, the 
regulatory specifications for safe reactor operations for Japan and United States 
require, at minimum, an accuracy of 2\%. In our study, a conservative uncertainty value of 2\% is adopted, including the error for 
thermal 
$LF$. We sample the thermal power of each core for every signal calculation.

The IBD cross section is extracted with a Monte Carlo sampling for each energy value at which the integrand of 
\equatname{\ref{eq:sigma_IBD}} is computed, where the adopted energy bin is equal to 1 keV.  

The global uncertainty of the reactor signal is evaluated by extracting simultaneously all the ingredients entering the 
uncertainty 
propagation procedure. This analysis is performed for 14 peculiar locations in the world, corresponding to sites hosting experiments that 
are currently ongoing or entering operation, as well as candidate sites for future neutrino experiments. Results are reported in 
\tabname{\ref{tab:15_sites}}, where the central values correspond to the medians and the errors are expressed as 1$\sigma$ uncertainties.

\subsection{Effect of long-lived isotopes}

During the operation of a nuclear reactor unstable fission fragments are constantly being produced, with half-lives in a wide range, 
from fractions of seconds up to 10$^{18}$ years. 
The Long-Lived Isotopes (LLIs) accumulate during the running of the reactor and consequently there exist off-equilibrium 
effects in the antineutrino spectrum from an operating reactor.
The $^{235}$U, $^{239}$Pu, and $^{241}$Pu antineutrino reference spectra entering the calculation of the total reactor spectrum are 
determined from the beta spectra measured after an exposure time to thermal neutrons of 12 hours ($^{235}$U) 
\cite{schreckenbach_determination_1985}, 1.5 days ($^{239}$Pu) \cite{von_feilitzsch_experimental_1982} and 1.8 days ($^{241}$Pu) 
\cite{hahn_antineutrino_1989}, which implies that long-lived fission fragments have not yet reached equilibrium.
Among unstable fission products of energy in the region $E_{\bar{\nu}}^{max}>1.806$ MeV, the most important LLIs having half-lives longer 
than 10 hours contribute only in the LER (see \tabname{\ref{tab:SNF}}), as the amplitude of the positive deviation from the reference 
spectra becomes negligible above 3.5 MeV \cite{mueller_improved_2011}.
The list of LLIs includes the Spent Nuclear Fuels (SNFs), i.e., $^{106}$Ru, $^{144}$Ce and $^{90}$Sr, having $\tau_{1/2} \sim$ yr.
As the off-equilibrium effects associated with the LLIs affect the antineutrino signal in the LER, understanding the LLIs contribution is a 
relevant issue in the geoneutrino framework.

We adopt the off-equilibrium corrections to the reference spectra reported in Table VII of \citet{mueller_improved_2011} in order to 
estimate the systematic uncertainty on the antineutrino signal due to the accumulation of LLIs during the running of the reactor.
As the operational run of a reactor usually lasts 1 year, signal values reported in \tabname{\ref{tab:15_sites}} include the 
300 days off-equilibrium correction to the reference $^{235}$U, $^{239}$Pu and $^{241}$Pu spectra published in 
\citet{mueller_improved_2011}.
After a time lapse on the order of one month with respect to the end of a reactor operating cycle, the SNF that has been pull 
out from the reactor contributes to approximately 0.6\% of the IBD unoscillated event rate in the Full Energy Region (FER) (see Fig.6 of 
\citet{bin_study_2012}). Each reactor is generally subject to a scheduled preventive maintenance on a yearly basis during which one third 
of the burnt fuel is typically transferred to the water pool located near the reactor core for cooling and shielding. As the exhausted 
fuel storage time can be as long as 10 years, the presence of the SNFs in the water pools can affect the reactor signal predictions, 
especially in the LER.
We estimate an average SNF half-life weighting the individual half-lives of the SNF species for the relative yields and fission fractions 
associated with each fissioning isotope, as stated in the following equation:
\begin{align}\label{eq:tSNF}
\tau_{1/2}^{SNF}&=\sum_{i= ^{144}\text{Ce},^{106}\text{Ru},^{90}\text{Sr}}k_{i}\tau_{1/2}^{i}\text{ ,}\nonumber 
\\&k_{i}=\sum_{l=^{235}\text{U},^{239}\text{Pu}}f_{l}\overline{Y_{l}^{i}}
\end{align}
where $\tau_{1/2}^{i}$ is the half-life of the $i$-$th$ SNF species, $f_{l}$ is the fission fraction (normalized to unity) for the $l$-$th$ 
fissioning isotope and $\overline{Y_{l}^{i}}$ is the production yield, with the normalization constraint $\sum_{l}\overline{Y_{l}^{i}}=1$.
Following this approach we estimate a SNF global half-life of 
\begin{equation}
 \tau_{1/2}^{SNF}=1.9 \text{ yr}
\end{equation}

The enhancement of the unoscillated IBD event rate due to the SNFs 
in the FER $\Delta N_{IBD}^{SNF}$ can be determined for a storage time T (expressed in units of 
years) according to 
\equatname{\ref{eq:S_SNF}}:
\begin{equation}\label{eq:S_SNF}
 \Delta N_{IBD}^{SNF}=\sum_{n=0}^{T} 0.2 \cdot \text{exp}\left(-\frac{n}{\tau^{SNF}}\right) 
\end{equation}
where we assume that every year a SNF mass equal to 1/3 of the reactor mass, decaying with a mean lifetime $\tau^{SNF}= 
\tau_{1/2}^{SNF}/ln(2)$, is transferred 
to the cooling pools. 

With the hypothesis of a 10 years storage time of SNFs, corresponding to the convergence of the series in \equatname{\ref{eq:S_SNF}}, we 
estimate a 2.4\% increase of the unoscillated IBD event rate in the LER, in agreement with \cite{gando_reactor_2013} and 
\cite{araki_experimental_2005}.
This potentially critical systematic uncertainty in geoneutrino measurements is not included in \tabname{\ref{tab:15_sites}}.

\subsection{Research reactors}

The research reactor (RR) class embraces a wide range of civil nuclear reactors that are generally not employed for power generation but 
they are mainly used as neutron sources, as well as for innovative nuclear energy researches and for teaching/training purposes. Among the 
major applications of the produced neutron beams are the non destructive tests of materials, neutron scattering experiments and the 
production of radioisotopes both for medical and industrial uses. 

According to the 2013 IAEA data published in \cite{research_reactor}, there are 247 operational RRs in the world accounting 
for a total thermal power of 2.2 GW, to be compared with the 1160 GW global thermal power generated by the 441 operational commercial 
reactors. Half of the RR thermal power is generated by only 8 reactors having an individual thermal power between 100 and 
250 MW. We calculate the expected reactor signal in the 14 experimental sites listed in \tabname{\ref{tab:15_sites}}, originating from the 
40 RRs that account for the 90\% of the thermal power considering an average 80\% annual load factor. 
The effect of this contribution is in any case smaller than 0.2\%, which can be considered as an 
upper limit enhancement of the commercial reactor signal.

\section{\label{results}Results and comments}
\begin{table*}
\caption{\label{tab:uncertainties}Uncertainty on the reactor signal in the FER for the long baseline experiments KL, BX and 
SNO+ due to the uncertainties on single inputs. Results are obtained by applying a Monte Carlo sampling of 
the input quantities according to the corresponding Probability Density Function (PDF). }
\begin{ruledtabular}
\renewcommand{\arraystretch}{1.4}
\begin{tabular}{
>{\centering\arraybackslash} m {0.8in} 
>{\centering\arraybackslash} m {0.3in} 
>{\centering\arraybackslash} m{0.4in}  
>{\centering\arraybackslash} m{0.8in} 
>{\centering\arraybackslash} m{0.7in} 
>{\centering\arraybackslash} m{1.5in}|
>{\centering\arraybackslash} m{0.7in}  
>{\centering\arraybackslash} m{0.7in} 
>{\centering\arraybackslash} m {0.7in} 
}
 & & & & & &\multicolumn{3}{c}{1$\sigma$ unc. on signal in the FER [\%]} \\
Input quantity&\multicolumn{2}{c}{Symbol}&PDF&1$\sigma$ unc. on input [\%]& Reference for input& BX&KL&SNO+ \bigstrut\\
\hline
\multirow{3}{\linewidth}{\parbox{\linewidth}{$\bar{\nu}$ oscillation}}& 
\multicolumn{2}{c}{$\delta m^{2}$}&Gaussian&3.4& \multirow{3}{1.7in}{\parbox{1.7in}{\citet{capozzi_status_2014}}}& 
$<0.1$&0.9&$<0.1$\bigstrut[t]\\
&\multicolumn{2}{c}{$sin^{2}(\theta_{12})$}& Gaussian& 5.5& & +2.4/-2.2&+2.1/-2.0&+2.4/-2.2\\
&\multicolumn{2}{c}{$sin^{2}(\theta_{13})$}& Gaussian& 8.5 & & 0.4&0.4&0.4 \\ 

\hline
\multirow{4}{\linewidth}{\parbox{\linewidth}{Energy released per fission}}&\multirow{4}{0.3in}{\parbox{0.2in}{$Q_{k}$}}& 
$Q_{^{235}U}$& \multirow{4}{0.7in}{\parbox{0.7in}{Gaussian}}& 0.1& \multirow { 4 } {
1.7in}{\parbox{1.7in}{\citet{ma_improved_2013}}}&
\multirow{4}{0.7in}{\parbox{0.7in}{$<0.1$}}&\multirow{4}{0.7in}{\parbox{0.7in}{$<0.1$}}&\multirow{4}{0.7in}{\parbox{0.7in}{$<0.1$}}\\

& &$Q_{^{238}U}$& &0.3& & & & \\

& &$Q_{^{239}Pu}$& &0.2& & & & \\

& &$Q_{^{241}Pu}$& &0.2& & & & \\
\hline

\multirow{4}{\linewidth}{\parbox{\linewidth}{Fuel composition}}&\multirow{4}{0.3in}{\parbox{0.2in}{$f_{k}$}}& 
$f_{^{235}U}$& \multirow{4}{0.7in}{\parbox{0.7in}{Constant PMF}}& \multirow{4}{0.7in}{\parbox{0.7in}{/}}& \multirow{4}
{ 1.7in}{\parbox{1.7in}{\tabname{\ref{tab:fission_fraction}}}}&
\multirow{4}{0.7in}{\parbox{0.7in}{0.1}}&\multirow{4}{0.7in}{\parbox{0.7in}{0.5}}&\multirow{4}{0.7in}{\parbox{0.7in}{$<0.1$}}\\

& &$f_{^{238}U}$& & & & & & \\

& &$f_{^{239}Pu}$& & & & & & \\

& &$f_{^{241}Pu}$& & & & & & \\

\hline
Thermal Power & \multicolumn{2}{c}{$P_{th}$}& Gaussian&2&\citet{djurcic_uncertainties_2009} &0.2&0.9&0.3\\
\hline
IBD cross
section&\multicolumn{2}{c}{$\sigma_{IBD}(E_{\bar{\nu}})$}&Gaussian&0.4&\citet{strumia_precise_2003}&$<0.1$&$<0.1$&$<0.1$\\
\end{tabular}
\end{ruledtabular}
\end{table*}

\begin{figure*}
\includegraphics[width=\textwidth]{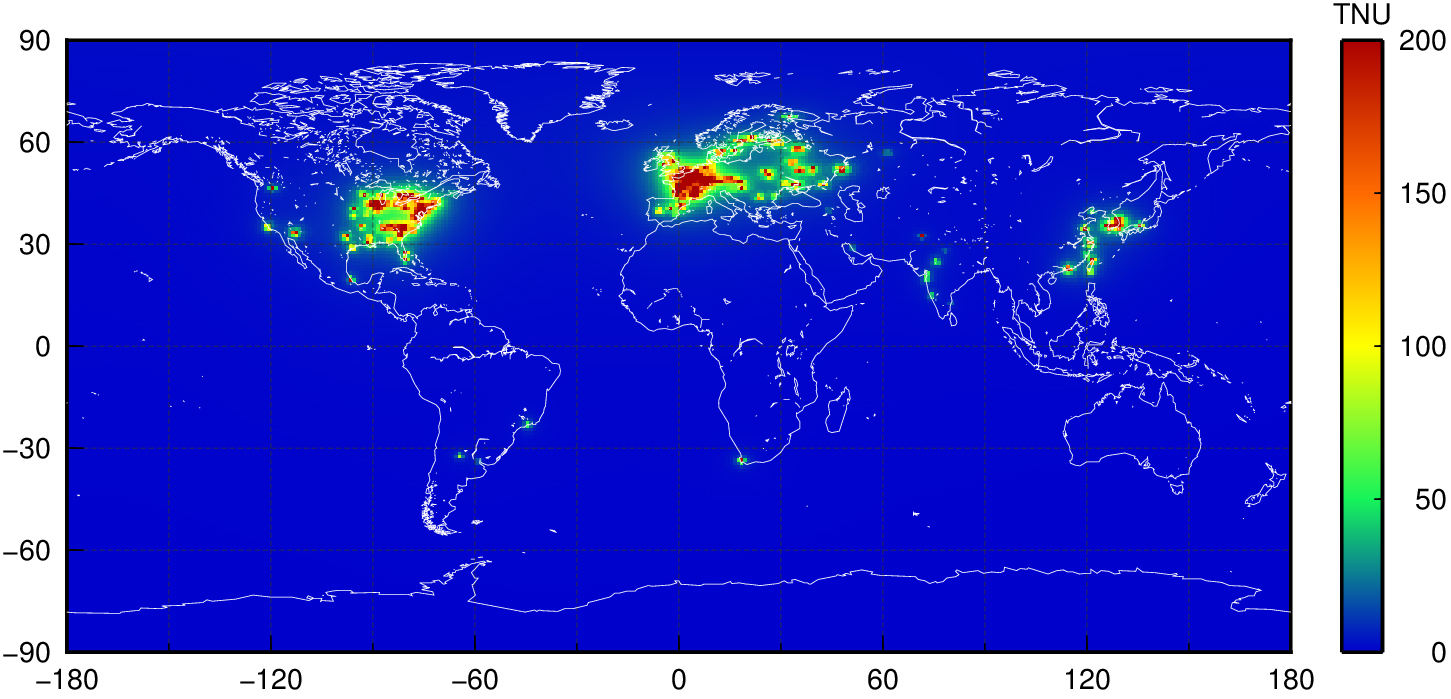}
\caption{\label{mappa}Map of the worldwide predicted antineutrino signals from nuclear power plants in the LER, 
expressed in TNUs. The 
map has a spatial resolution of 1° x 1° and it is produced with 2013 operational data on nuclear power plants.}
\end{figure*}

\begin{table*}
\caption{\label{tab:15_sites}Predicted antineutrino signals (in TNU) from nuclear power plants in the FER (R$_{\text{FER}}$) and in the 
LER (R$_{\text{LER}}$) obtained with 2013 reactor operational data, together with the expected geoneutrino signals (G) and 
R$_{\text{LER}}$/G ratios at current and proposed neutrino experimental sites. Antineutrino signals in the FER and in the LER include 
the off-equilibrium contribution due to the accumulation of the LLIs during the running of the reactor. For the KL experiment we report 
also the values obtained using 2006 reactor operating records. For the Juno experiment we predict the 2020 reactor signals, considering as 
operating with a 80\% 
annual average load factor the Yangjiang (17.4 GW) and Taishan (18.4 GW) nuclear power stations which are actually under construction.}

\begin{ruledtabular}

\renewcommand{\arraystretch}{1.7}
\begin{tabular}{c c c c c c c}
Site&Experiment&Coordinates&G [TNU]& R$_{\text{FER}}$ [TNU]&R$_{\text{LER}}$ [TNU]&R$_{\text{LER}}$/G \bigstrut \\
\hline
Gran Sasso (IT)\footnote{IT: Italy, JP: Japan, CA: Canada, CH: China, SK: South Korea, US: United States of America, FI: Finland, 
UK: United Kindom, SP: Spain, FR: France, RO: Romania, PL: Poland, RU: Russia.}&Borexino&42.45 N, 13.57 
E\footnote{\label{foot:huang}\citet{huang_reference_2013}}& 40.3$^{+7.3}_{-5.8}$  &83.3 $^{+2.0}_{-1.9}$&22.2$^{+0.6}_{-0.6}$ &0.6 
\bigstrut[t] \\
Sudbury (CA)&SNO+&46.47 N, 81.20 W\textsuperscript{\ref{foot:huang}}& 45.4$^{+7.5}_{-6.3}$& 190.9$^{+4.6}_{-4.2}$&47.8$^{+1.7}_{-1.4}$ &  
1.1 \\

\multirow{2}{0.15\linewidth}{\parbox{\linewidth}{Kamioka (JP)}}& 
\multicolumn{1}{c}{\multirow{2}{0.15\linewidth}{\parbox{\linewidth}{KamLAND}}}&\multicolumn{1}{c}{\multirow{2}{0.15\linewidth}{\parbox
{\linewidth} { 36.43 N, 137.31 
E\textsuperscript{\ref{foot:huang}}}}}& \multicolumn{1}{c}{\multirow{2}{0.15\linewidth}{\parbox{\linewidth}{31.5$^{+4.9}_{-4.1}$}}}&65.3$^{
+1.7}_{-1.6}$&18.3$^{+0.6}_{-1.0} $& 0.6\\

& & & &625.9$^{+14.5}_{-13.2}$\footnote{\label{foot:2006}2006 reactor operational 
data.}&168.5$^{+5.7}_{-6.3}$\textsuperscript{\ref{foot:2006}}&5.3\textsuperscript{\ref{foot:2006}}\\

\multirow{2}{0.15\linewidth}{\parbox{\linewidth}{DongKeng (CH)}}& 
\multicolumn{1}{c}{\multirow{2}{0.15\linewidth}{\parbox{\linewidth}{Juno}}}&\multicolumn{1}{c}{\multirow{2}{0.15\linewidth}{\parbox
{\linewidth} {22.12 N, 112.52 E\footnote{\label{foot:ciuffoli}\citet{ciuffoli_advantages_2014}}}}}  & 
\multicolumn{1}{c}{\multirow{2}{0.15\linewidth}{\parbox{\linewidth}{39.7$^{+6.5}_{-5.2}$}}}&95.3$^{+2.6}_{-2.4}$ &26.0$^{+2.2}_{-2.3}$& 
0.7\\

& & & &1566$^{+111}_{-100}$\footnote{\label{foot:2020}2013 reactor operational data plus Yangjiang (17.4 GW) and Taishan (18.4 GW) nuclear 
power stations operating with a 80\% average annual load 
factor.}&354.5$^{+44.5}_{-40.6}$\textsuperscript{\ref{foot:2020}}&8.9\textsuperscript{\ref{foot:2020}}\\

GuemSeong (SK)&RENO-50&35.05 N, 126.70 E\textsuperscript{\ref{foot:ciuffoli}}&
38.3$^{+6.1}_{-4.9}$& 1128$^{+75}_{-67}$&178.4$^{+20.8}_{-19.6}$ & 4.7\\

Hawaii (US)&Hanohano&19.72 N, 156.32 W\textsuperscript{\ref{foot:huang}}& 12.0$^{+0.7}_{-0.6}$&3.4$^{+0.1}_{-0.1}$&0.9$^{+0.02}_{-0.02}$ & 
0.1\\
Pyh\"{a}salmi (FI)&LENA&63.66 N, 26.05 E\textsuperscript{\ref{foot:huang}}& 45.5$^{+6.9}_{-5.9}$&66.1$^{+1.6}_{-1.5}$&17.0$^{+0.5}_{-0.4}$ 
& 0.4\\

Boulby (UK)&LENA&54.55 N, 0.82 W\textsuperscript{\ref{foot:huang}}& 39.2$^{+6.3}_{-4.9}$& 1234$^{+35}_{-35}$&240.6$^{+11.5}_{-11.9}$ &6.1 \\

Canfranc (SP)&LENA&42.70 N, 0.52 W\textsuperscript{\ref{foot:huang}}&40.0$^{+6.4}_{-5.1}$&247.4$^{+5.8}_{-5.5}$&70.3$^{+1.6}_{-1.7}$ & 
1.8\\

Fréjus (FR)&LENA&45.13 N, 6.68 E\textsuperscript{\ref{foot:huang}}& 42.8$^{+7.6}_{-6.4}$&546.7$^{+11.9}_{-10.5}$&126.0$^{+5.4}_{-5.1}$ & 
2.9\\

Sl\u{a}nic (RO)&LENA&45.23 N, 25.94 E\textsuperscript{\ref{foot:huang}}&45.1$^{+7.8}_{-6.3}$& 109.2$^{+2.7}_{-2.5}$&29.6$^{+0.7}_{-0.7}$ 
&0.7\\

Sieroszowice (PL)&LENA&51.55 N, 16.03 E\textsuperscript{\ref{foot:huang}}&43.4$^{+7.0}_{-5.6}$& 153.3$^{+3.8}_{-3.6}$&41.4$^{+1.1}_{-1.1}$ 
&1.0\\

Homestake (US)&/&44.35 N, 103.75 W\textsuperscript{\ref{foot:huang}}&48.7$^{+8.3}_{-6.9}$& 30.4$^{+0.7}_{-0.7}$&8.0$^{+0.2}_{-0.2}$ &0.2\\

Baksan (RU)&/&43.20 N, 42.72 E\textsuperscript{\ref{foot:huang}}&47.2$^{+7.7}_{-6.4}$&36.6$^{+0.9}_{-0.8}$&9.6$^{+0.3}_{-0.3}$ 
&0.2\bigstrut[b]\\
\end{tabular}
\end{ruledtabular}
\end{table*}

In \tabname{\ref{tab:uncertainties} we report the results of our estimate of the uncertainties on 
the reactor signal due to the $1\sigma$ errors associated with single inputs. For the three operative long baseline 
experiments 
the major effect is attributed to $sin^{2}(\theta_{12})$, which generates an uncertainty on the signal of approximately 2.2\% at 
1$\sigma$ level.

The impact on the signal uncertainty due to the uncertainties on reactors 
thermal power and on the fission fractions is highly site dependent.  It emerges as a combined effect of the different reactor 
distances from the experimental sites and of the number and class of close-by reactors.

In 2013, approximately 60\% of the signal predicted at KL is almost equally shared between just two Japanese reactor cores (Ohi stations 3 
and 4) which are located 180 km far away from the Kamioka mine. 
The same signal percentage is produced at BX by approximately 60 reactors located within a radius of 1000 km, where each core 
contributes to less than 3\% of the signal. With respect to SNO+, 20 cores situated within a 500 km radius from the experimental site 
provide approximately 60\% of the signal, each core contributing to 6\% of the signal at maximum (see \figname{\ref{contributi}}).
As a consequence, the uncertainty on reactors thermal power generates at KL an uncertainty on the signal three times higher 
than what estimated for BX and SNO+, on the order of 1\%.

Fission fractions give rise to a few tenths of percent $1\sigma$ uncertainty on the reactor signal. 
The effect of fission fractions at KL is five times larger with respect to what estimated at BX: this behavoiur reproduces the 
one already observed for the thermal power, and is also related to the fact that reactors giving the highest 
contributions to the signal belong to the same reactor class. On the other hand, SNO+ is almost insensitive to fission fractions 
variability, since the signal is dominated by the Canadian PHWRs, for which a fixed single set of power fractions is currently available.

In this work we present also a worldwide map (with a 1° x 1° spatial resolution) of expected reactor signals in the LER 
expressed in TNUs, produced using 2013 operational reactor data (see \figname{\ref{mappa}}). This map provides evidence regarding the sites 
demonstrating the best discrimination power on geoneutrino measurements. 

A particular focus is dedicated to sites hosting ongoing neutrino experiments (KL and BX), experiments 
entering operation (SNO+), and candidate sites for future experiments (Juno, RENO-50, Hanohano, LENA, Homestake, Baksan). For these 
specific locations we evaluate the expected reactor signal both in the FER (R$_{\text{FER}}$) and in the LER (R$_{\text{LER}}$) and the 
predicted geoneutrino signal G on the base of the reference Earth model published by \citet{huang_reference_2013}. We also evaluate the 
ratio R$_{\text{LER}}$/G, which can be considered as a figure of merit for assessing the sensitivity to geoneutrinos at a given site (see 
\tabname{\ref{tab:15_sites}}). 

The reactor signals R$_{\text{FER}}$ and R$_{\text{LER}}$ are determined as median values of the signal distributions obtained from the 
Monte Carlo 
calculation. For each site the signals are computed 10$^{4}$ times using the \citet{mueller_improved_2011} analytical parametrization of 
the reactor spectrum, including the 300 days off-equilibrium correction due to the LLIs, and simultaneously extracting, according to 
the corresponding PDF, all the inputs entering the uncertainty propagation procedure as described in \secname{\ref{sec:uncertainties}}.
For the long baseline experiments, signal errors are evaluated as 1$\sigma$ uncertainties and are estimated to be on the order of 3\% and 
4\% for the signal in the FER and in the LER, respectively. 
Ratios R$_{\text{LER}}$/G between predicted geoneutrino and reactor signals in the LER (calculated using 2013 reactor operational features) 
show the high discrimination power on geoneutrinos achievable at Hawaii (R$_{\text{LER}}$/G = 0.1), Homestake and Baksan 
(R$_{\text{LER}}$/G = 0.2).
In 2013 a relatively high sensitivity to geoneutrinos is attainable at Kamioka (R$_{\text{LER}}$/G  = 0.6) thanks to the protracted 
shutdown of the Japanese reactors after the Fukushima accident, in comparison with the much lower geoneutrino discrimination power 
of 2006 (R$_{\text{LER}}$/G  = 5.4) when the Japanese power industry was fully operational.
Moreover, Juno appears to be a good candidate site for geoneutrino measurements according to 2013 reactors operating 
status. If the experiment construction is achieved before the completion of the Yangjiang (17.4 GW) and Taishan (18.4 GW) nuclear power 
plants, the 20 kton detector will reach a 10\% accuracy on geoneutrinos in approximately 105 days (assuming a C$_{17}$H$_{28}$ 
liquid scintillator composition, a 100\% detection efficiency and that the geoneutrino background is due only to reactor antineutrinos). In 
contrast, we predict that the ratio R$_{\text{LER}}$/G dramatically 
increases from 0.7 to 8.9 if we consider both Chinese power stations to operate with an annual average load factor of 80\%.

\begin{table*}
\caption{\label{tab:models}Reactor signals (without the LLIs contribution) in the FER and in the LER obtained with the analytical 
parametrization of the 
reactor spectra from \citet{huber_determination_2011}, \citet{huber_precision_2004}, 
\citet{vogel_neutrino_1989} and \citet{mueller_improved_2011} for the BX, KL and SNO+ experiments. Since in 
\cite{huber_precision_2004} and \cite{huber_determination_2011} there is no 
analytical expression for the $^{238}$U antineutrino spectrum, the one reported in \citet{mueller_improved_2011} is used in these two 
cases.}
\begin{ruledtabular}
\renewcommand{\arraystretch}{1.8}
\begin{tabular}{c c c c c c c}
 & \multicolumn{3}{c}{R$_{\text{FER}}$ [TNU]}&\multicolumn{3}{c}{R$_{\text{LER}}$ [TNU]}\\
Reactor spectra model&BX&KL&SNO+&BX&KL&SNO+  \\
\hline
\citet{mueller_improved_2011}& 83.2 $^{+2.0}_{-1.8}$& 65.3 $^{+1.7}_{-1.6}$&190.2 $^{+4.8}_{-4.3}$ & 22.1$^{+0.6}_{-0.5}$ & 
18.3$^{+0.6}_{-1.0}$&47.2 $^{+1.7}_{-1.4}$\\
\citet{huber_determination_2011}+$^{238}$U \citet{mueller_improved_2011}&83.9 $^{+2.0}_{-1.8}$&65.9$^{+1.7}_{-1.6}$ &192.0 $^{+4.9}_{-4.3}$ 
& 22.0$^{+0.6}_{-0.5}$&18.3$^{+0.6}_{-1.0}$&47.1 $^{+1.7}_{-1.4}$\\
\citet{huber_precision_2004}+$^{238}$U \citet{mueller_improved_2011}&81.2 $^{+2.0}_{-1.8}$&63.7$^{+1.6}_{-1.5}$ & 185.5 $^{+4.7}_{-4.1}$& 
21.7$^{+0.6}_{-0.5}$&18.0$^{+0.6}_{-1.0}$&46.3 $^{+1.7}_{-1.4}$\\
\citet{vogel_neutrino_1989}&81.6 $^{+2.0}_{-1.8}$&63.9$^{+1.6}_{-1.6}$ &187.1 $^{+4.7}_{-4.2}$ & 
21.6$^{+0.5}_{-0.6}$&17.9$^{+0.6}_{-1.0}$&46.0 $^{+1.7}_{-1.4}$\\
\end{tabular}
\end{ruledtabular}
\end{table*}

\begin{figure*}
\includegraphics[width=0.8\linewidth]{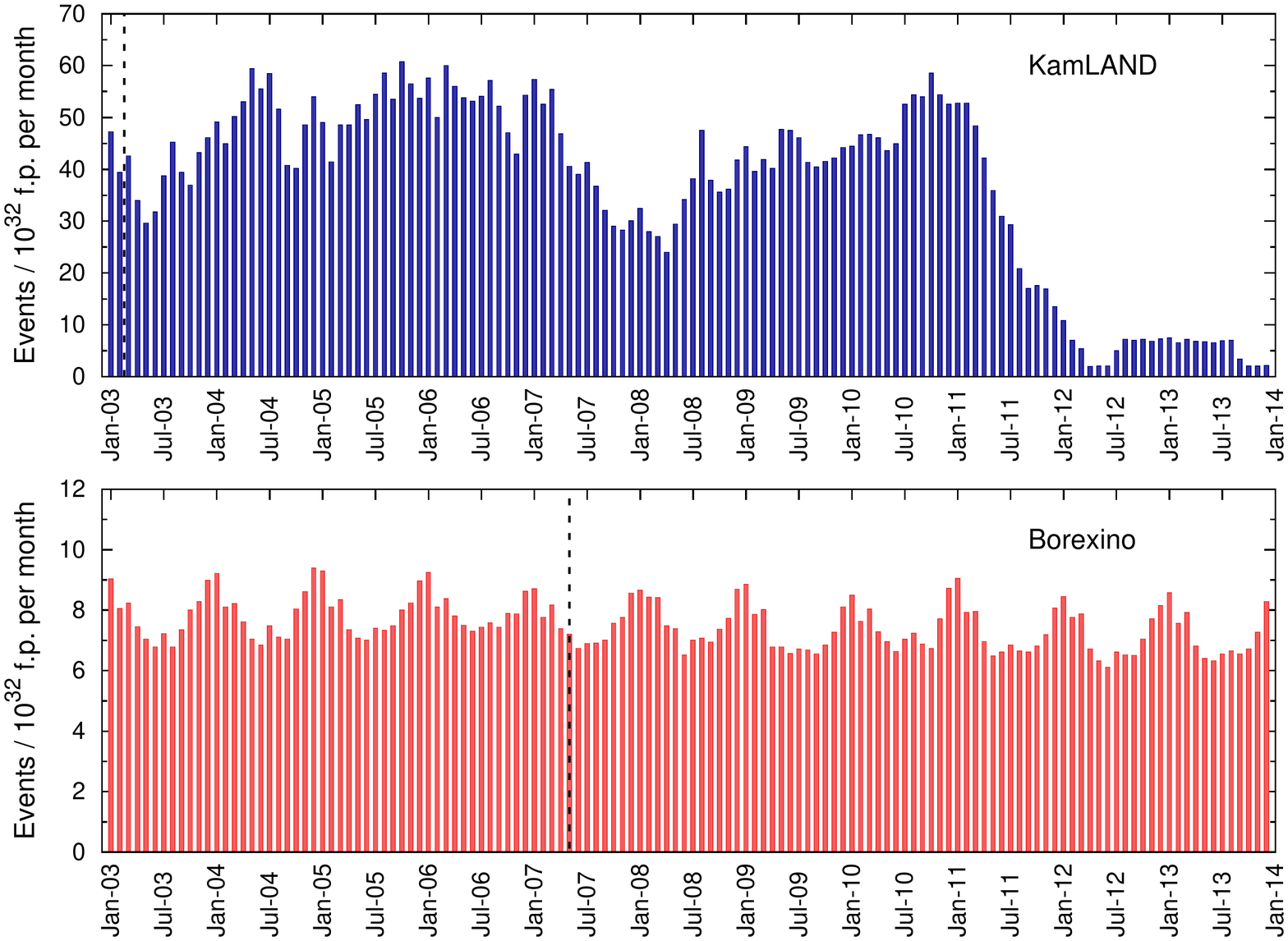}
\caption{\label{eventi}Reactor signals in the FER for the KL experiment (blue panel) and for the BX experiment 
(red panel), calculated from January 2003 to December 2013 on a monthly timeline. The vertical dashed lines indicate the 
data taking start of the experiments (March 2003 for KL and May 2007 for BX).}
\end{figure*}

\begin{figure*}
\includegraphics[width=0.8\linewidth]{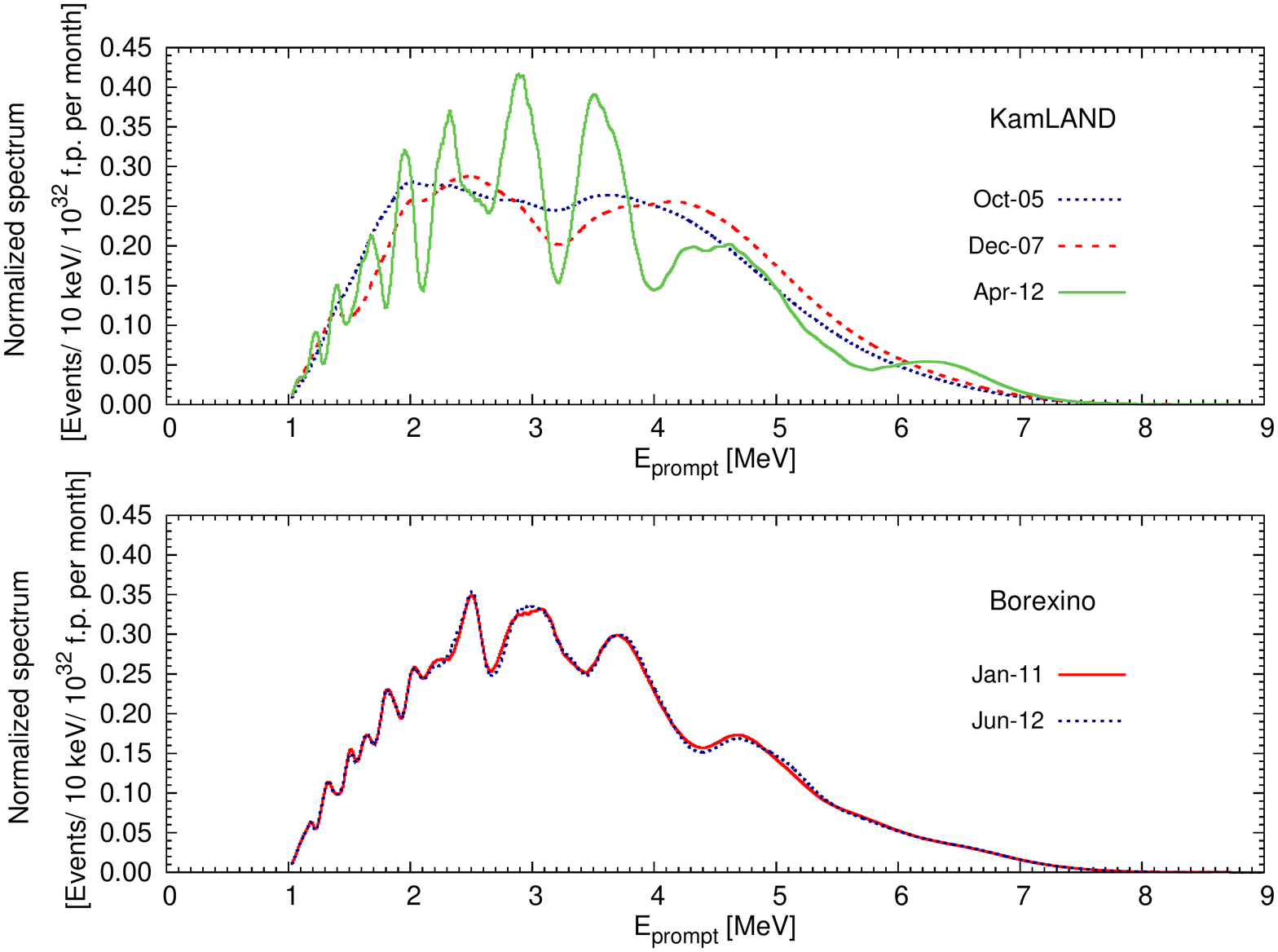}
\caption{\label{spettri}Reactor antineutrino spectra above IBD threshold for the KL experiment (upper panel) and for the BX 
experiment (lower panel) calculated over different data taking periods. KL spectra are evaluated over three 
peculiar time intervals, corresponding to a maximum, an average and a minimum expected reactor signal (October 2005, December 2007 and April 
2012, respectively). BX spectra are calculated in correspondence to a winter and a summer seasonal signal variation (January 2011, June 
2012). All the spectra are normalized to the signal corresponding to the specific month.}
\end{figure*}

\begin{figure*}
\includegraphics[width=0.7\linewidth]{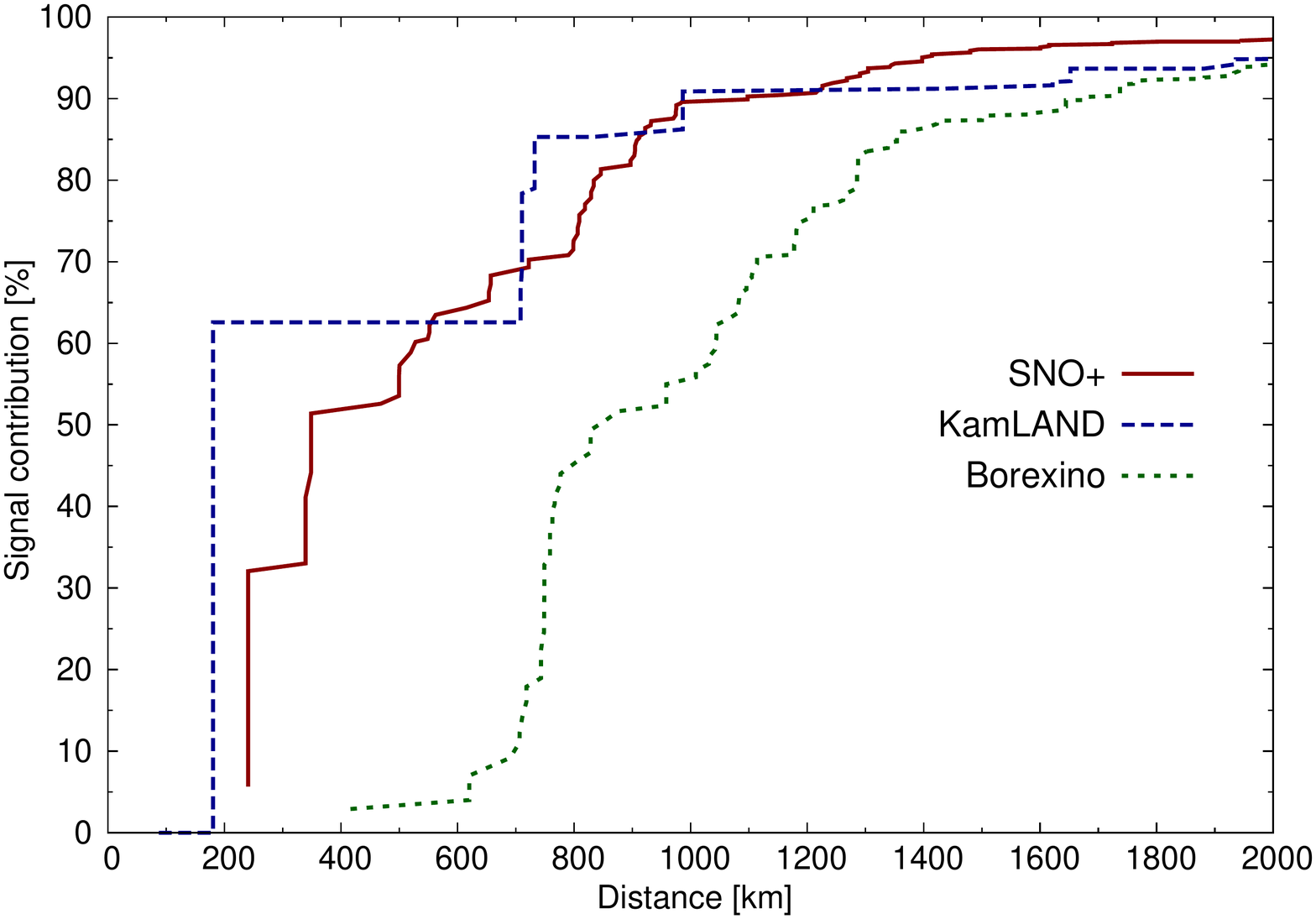}
\caption{\label{contributi}Cumulative percentage contribution to the total expected reactor signal as function of the 
distance of the reactors from the experimental site for KL, BX and SNO+. Data refer to 2013 reactor operational 
period.}
\end{figure*}

\begin{figure*}
\includegraphics[width=\textwidth]{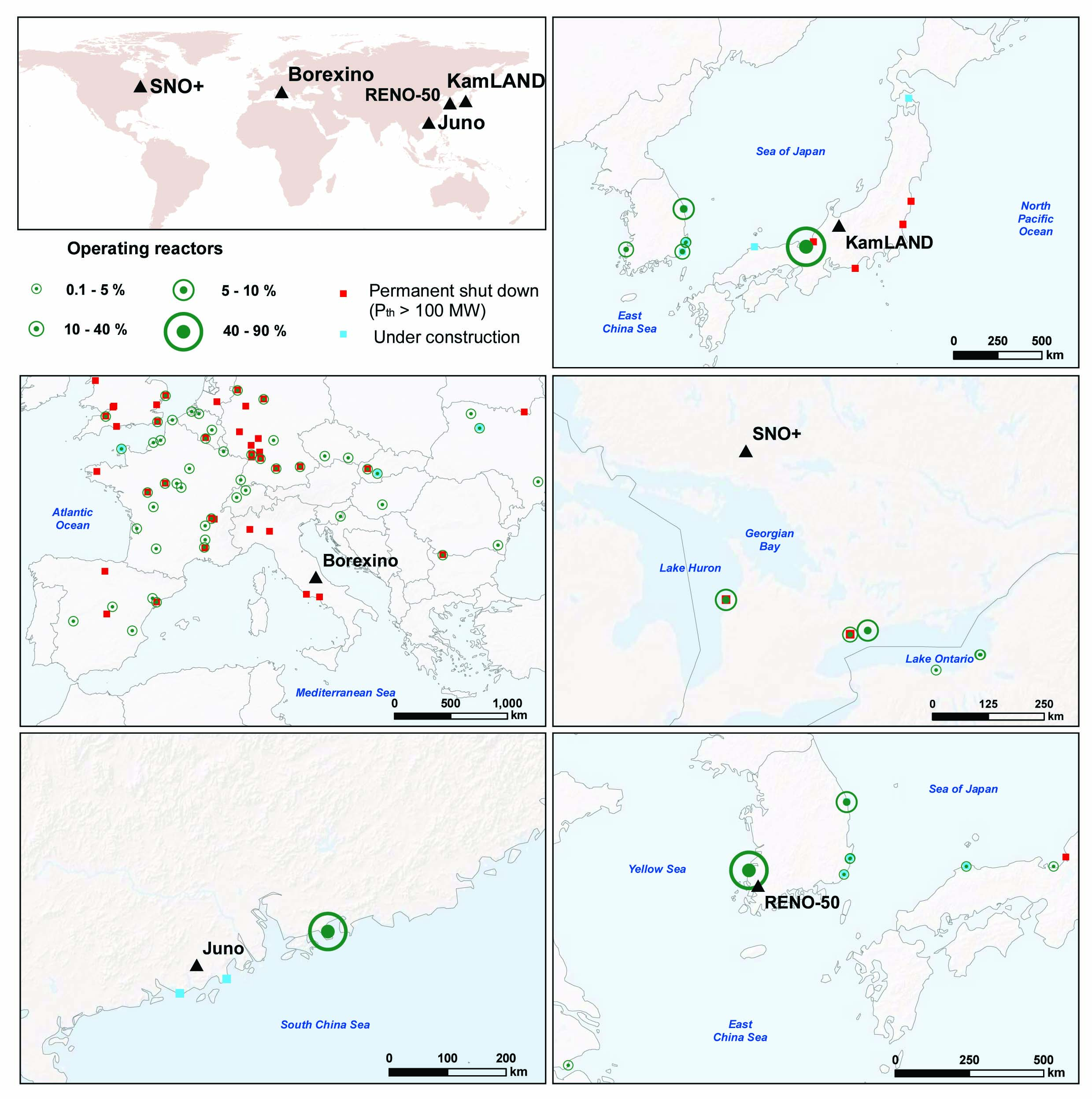}
\caption{\label{fig_mappe_new}Location map of the percentage contributions to the reactor signal given by the close-by reactors 
for the 
three long baseline experiments KL, BX and SNO+ and for the proposed medium baseline experiments Juno and RENO-50. The map is 
produced with 2013 reactor operational data.}
\end{figure*}

To estimate the variability in the expected reactor signal due to different reactor spectra, we calculate the predicted signals at 
KL, BX and SNO+ using three alternative parametrizations of the antineutrino spectra, i.e. the ones published 
by \citet{huber_determination_2011}, \citet{huber_precision_2004} and \citet{vogel_neutrino_1989} (see \tabname{\ref{tab:models}}). There 
is no expression for the $^{238}$U spectrum in \cite{huber_precision_2004} and \cite{huber_determination_2011}, as these 
parametrizations are based on the conversion of ILL beta spectra. Therefore, for these two sets of spectra, the 
adopted functional expression for the $^{238}$U antineutrino spectrum is provided by \citet{mueller_improved_2011}. Median signal 
values are shown in \tabname{\ref{tab:models}, together with the 1$\sigma$ uncertainties evaluated via Monte Carlo 
sampling.
The maximum signal spread associated with the employment of different analytical functions as phenomenological parametrization of the 
reactor antineutrino spectrum is of the same order as the global uncertainty on the signal resulting from the combined effect of all the 
other input quantities. Therefore, the reactor antineutrino spectrum emerges as the most critical component in the signal calculation.

We present a time profile of the expected reactor antineutrino signals at KL and BX over a period of 10 years on a monthly time-line, from 
2003, when the KL detector entered operation, to 2013 (see \figname{\ref{eventi}}).

The BX time profile exhibits a seasonal variation, suggesting that the periodic signal shape could be possibly 
implemented in the event analysis. The highest and lowest reactor signals occur respectively in correspondence with the 
cold and warm seasons, being the electricity demand typically higher during the winter. In 
connection to this, it can be noticed that refueling and maintenance for nuclear units are typically performed in the spring and fall 
seasons, when demand for electricity is generally lower. In \cite{borexino_collaboration_measurement_2013} the antineutrino event analysis 
on a 1353 days data taking period shows a good agreeement with our prediction, although the seasonal variation has been not studied.

The KL signal time profile is instead highly affected by the operating conditions of the Japanese reactors. The shutdown of nuclear power 
plants concomitant to strong earthquakes in Japan is therefore manifestly visible as a pronounced decrease in the evaluated reactor signal. 
In particular, there is clear evidence of the protected shutdown of the Kashiwazaki-Kariwa and Hamaoka nuclear power plants subsequent to 
the Chūetsu earthquake in July 2007 and of the protected shutdown of the entire Japanese nuclear reactor industry following the Fukushima 
nuclear accident in March 2011.

The different signal time profiles for the two experiments reflect also in different reactor antineutrino spectra 
(see \figname{\ref{spettri}}). As understood from the contribution on the signal uncertainty given by the reactor thermal power and fission 
fraction uncertainties (see \tabname{\ref{tab:uncertainties}}), the antineutrino spectrum at BX is relatively insensitive to 
different operational conditions of individual nuclear power plants, as there are no close-by reactors dominating the antineutrino flux. 
Conversely, detailed information on the operating status of the near reactors emerges as a fundamental piece of knowledge for modelling the 
reactor spectrum at KL.

The distribution of the cumulative percentage contribution to the total reactor signal as a function of the distance of the reactors from 
the experimental site (see \figname{\ref{contributi}}) yields a hint of the level of criticality associated with the knowledge of the 
operational parameters of reactors.
The KL distance profile has a step-like function shape: the first discontinuity is observed at 180 km where the signals 
coming from units 3 and 4 of the Ohi nuclear power plant sum up and provide approximately 60\% of the 
total reactor signal. The second and third discontinuities in the KL distribution (85\% and 90\% of the total signal, respectively) 
occur for a reactor-detector distance of 730 km (corresponding to the contribution given by all operating Japanese reactors and by the 
South Korean reactors located on the East coast) and 990 km (summing up the contribution of the Hanbit power plant, located in West South 
Korea). 
The BX distance profile is smoother compared to that of KL as the reactor signal is gradually spread out over the 
European countries. With respect to BX, the closest power station is at a distance of 415 km (Slovenia), which contributes the major 
fraction of the reactor signal, i.e., approximately 3\%. With respect to the SNO+ experiment, the distribution is dominated at short 
distance by the Canadian Bruce power station, corresponding to the first step in the distance profile at 240 km (32\% of the signal). The 
second step is associated with the Pickering and Darlington power plants and occurs at a site-reactor distance of 350 km (51\% of the 
signal). For a site-reactor distance greater than 500 km the profile levels out due to the contributions given by the more distant power 
stations located in the United States.

The percentage contributions to the signal given by the relatively close reactors at long baseline experiments (KL, BX and SNO+) and at 
proposed medium baseline experiments (Juno and RENO-50) are displayed on a location map (see \figname{\ref{fig_mappe_new}). In addition to 
the contributions of operating power plants in 2013, nuclear stations under construction are displayed. 

\section{\label{conclusions}Conclusions}
One of the primary goals of the current and proposed reactor neutrino experiments is to investigate the neutrino properties at different 
wavelengths 
according to different reactor-detector baselines. While shedding light on the oscillatory neutrino nature, neutrino experiments also 
provide insight into the Earth’s interior via the detection of geoneutrinos. In this framework, nuclear power plants emerge as the 
most 
severe background sources as approximately 27\% of the reactor event rate is recorded in the geoneutrino energy window.
The main results of this work are as follows.
\begin{itemize}
\item We evaluated the expected antineutrino signal from not movable reactors for 14 peculiar locations in the world,
estimating its uncertainties in view of reactors operational information yearly published by the Power Reactor Information System (PRIS) of 
the International Atomic
Energy Agency (IAEA). A comprehensive database concering nuclear power plants operational status is published at 
\url{www.fe.infn.it/antineutrino} and we plan to update it every year. We evaluated the expected antineutrino signal from reactors and from 
the Earth for 14 peculiar locations in the world, corresponding to sites hosting experiments that are currently ongoing or entering 
operation, as well as candidate sites for future neutrino experiments. 

\item The Monte Carlo method applied for the propagation of (uncorrelated) uncertainties on reactor signals associated with the input 
quantities provided an overall uncertainty for the long baseline experiments of approximately 3\% in the FER and of approximately 4\% 
in the LER, for a fixed analytical expression of the reactor spectrum. The reactor signal uncertainty is dominated by 
$sin^{2}(\theta_{12})$, which solely provides an uncertainty of approximately 2.2\% in the FER for KL, BX and SNO+.

\item 
We performed a comparison of the reactor signals obtained using different reactor spectra, which revealed that the uncertainty related 
to the antineutrino spectrum is as critical as the combined uncertainty of the other input quantities appearing in the signal calculation.

\item We discussed the effect of the systematic enhancement of the reactor antineutrino spectrum due both to the 
accumulation of the LLIs during the operation of a reactor and to the storage of the SNFs in the cooling pools. 
We estimate a 2.4\% increase of the unoscillated IBD event rate in the LER due to the SNFs that potentially can be a critical systematic 
uncertainty in geoneutrino measurements.

\item We estimated that the RRs producing a total thermal power of 2.2 GW contribute less than 0.2\% to the commercial reactor signal in 
the investigated 14 sites. 

\item We presented a multitemporal analysis of the expected reactor signal at BX and KL over a time lapse of 10 years. With 
respect to BX, a periodic seasonal signal variation associated with the lower fall-spring 
electricity demand 
is recognized: expected reactor signals are relatively insensitive to the operational conditions of single cores, since there 
are no close-by reactors dominating the antineutrino flux. Conversely, the KL signal time profile is governed by the Japanese nuclear 
industry operational status, which make the shutdown of nuclear power plants concomitant to strong earthquakes manifestly 
visible.

\end{itemize}

\section*{Acknowledgements}

The authors would like to thank C. Rott for the useful information on RENO-50 and J. Mandula for the valuable 
help in the compilation of the nuclear reactors database. We are also grateful to J. Esposito, T. Lasserre, D. Lhuillier, L. Ludhova, Th. A. 
Mueller and S. Zavatarelli for the useful comments. We appreciate geological insights from R. L. Rudnick, W. F. McDonough 
and Y. Huang. We wish to thank the two anonymous reviewers for their detailed and thoughtful reviews. 

This work was partially supported by 
the Istituto Nazionale di Fisica Nucleare (INFN) through the ITALRAD Project and by the 
University of Ferrara through the research initiative ``Fondo di Ateneo per la Ricerca scientifica FAR 2014''.

\section*{Appendix}

The database of the operating commercial reactors is compiled starting from 2003 up to now on a yearly basis and updated using 
the operational information yearly published by the Power Reactor Information System (PRIS) of the 
International Atomic Energy Agency (IAEA) (\url{http://www.iaea.org/pris/home.aspx}).
The nuclear power plant database contains 19 columns, structured as follows (for a given year of operation):
\begin{itemize}
 \item core country acronym;
 \item core name;
 \item core location (latitude and longitude in decimal degrees);
 \item core type;
 \item use of MOX (1 for yes, 0 for no);
 \item thermal power $P_{th}$ [MW];
 \item 12 columns listing the load factor for each month, expressed in percentage. 
\end{itemize}

Latitude and longitude of core locations are taken from the World Nuclear Association 
Database (\url{http://world-nuclear.org/NuclearDatabase/Default.aspx?id=27232}).
Core country acronyms, core name, core type, thermal and electrical power and load factors are defined and published in the PRIS annual 
publication entitled ``Operating Experience with Nuclear Power Stations in Member States''.

\nocite{*}

\bibliography{Exported}

\end{document}